\documentclass[aps,pra,showpacs,eqsecnum,twocolumn]{revtex4}
\usepackage{graphicx}
\usepackage{bm}
\begin{document}

\title{Coherence for vectorial waves and majorization}

\author{Alfredo Luis}
\affiliation{Departamento de \'{O}ptica, Facultad de Ciencias
F\'{\i}sicas, Universidad Complutense, 28040 Madrid, Spain}

\begin{abstract}
We show that majorization provides a powerful approach to the coherence conveyed by partially 
polarized transversal electromagnetic waves. Here we present the formalism, provide some 
examples and compare with standard measures of polarization and coherence of vectorial waves.
\end{abstract}

\pacs{42.25.Kb, 42.25.Ja, 42.25.Hz, 89.70.Cf, 02.50.$-$r}


\maketitle

\section{Introduction}

Coherence is a basic physical property that emerges in very different contexts, from classical 
optics to quantum mechanics. Therefore the efforts to find a proper measure of coherence 
seem justified, specially after its identification as a resource \cite{BCP}. 

In this work we focus on the assessment of the coherence conveyed by pairs of partially 
polarized electromagnetic beams in classical optics. Coherence has two extreme physical 
manifestations: interference, as coherence when superimposing beams with the same 
vibrational state, and polarization, as coherence when superimposing beams with 
orthogonal vibration states. This makes specially attractive the analysis of coherence in the 
superposition of partially polarized waves. The complexity of the subject has motivated the 
introduction of complementary approaches focusing on different perspectives \cite{KW, TSF, RG}. 
Among them, it makes sense to address the idea of global conference embracing both  
interference and polarization at once  \cite{OYK,LU}. Parallels can also be drawn to its quantum 
counterpart through the idea of state purity \cite{QvC1,QvC2,PP}.

Besides standard measures, entropies are in general good candidates to asses statistical 
properties. Entropy has already been used to measure polarization \cite{PE,QvC2}, being 
specially useful for situations beyond Gaussian or second-order statistics. A rather attractive 
feature when dealing with entropies is the emergence of majorization as a kind of meta-measure 
that establishes a once-for-all partial ordering of statistics \cite{MO}. This partial relation is 
respected by the Schur-concave functions, that include the Shannon and R\'{e}nyi entropies 
\cite{RB}.

We will study the coherence conveyed by two transversal  waves, this is four electric field 
components. This may be the case of the interference of partially polarized waves 
emerging from the pinholes of a Young interferometer. The statistics will be fully specified by 
their second-order field correlations in the space-frequency domain, this is a $4 \times 4$ 
positive cross-spectral density tensor $\bm{\Gamma}$. The entropies and majorization will be 
applied to its normalized spectrum, this is to the eigenvalues of  $\bm{\Gamma} /\mathrm{tr} 
\bm{\Gamma}$. As a proper precedent we can mention the application of majorization to 
the analysis of polarization of three-dimensional fields \cite{GJ}. The conclusions will be 
contrasted to other accounts of coherence more focused on the ideas of field correlation or 
fringe visibility. 

\section{Settings: Field states and coherence measures}

For definiteness let us focus on two vectorial electric fields $\bm{E}$ at two spatial points 
$\bm{r}_{1,2}$  with just two non vanishing components at each point, say $E_{x,y}$. This 
can be the transverse electric field at the two pinholes of a Young interferometer. The complete
system is made of four scalar electric fields that we will consider in the space-frequency domain 
$E_\ell \left ( \bm{r}_j , \omega  \right )$ with $j=1,2$, $\ell = x,y$, and the temporal frequency 
$\omega$ will  be omitted from now on. Their statistics will be completely accounted for by the 
second-order field correlations gathered by the cross-spectral density tensor,  that in our case 
is a $4 \times 4$ nonnegative matrix  $\bm{\Gamma}$, with
\begin{equation}
 \bm{\Gamma} = \pmatrix{\bm{\Gamma}_{1,1} & \bm{\Gamma}_{1,2} \cr \bm{\Gamma}_{2,1}  & 
 \bm{\Gamma}_{2,2}} ,
 \end{equation}
 where $\Gamma_{j,k}$ are $2 \times 2$ correlation 
matrices with  $\Gamma_{j,k} = \Gamma_{k,j}^\dagger$,
 \begin{equation}
 \Gamma_{j,k} = \left \langle \bm{E} \left  ( \bm{r}_j \right ) \bm{E}^\dagger \left  ( \bm{r}_k \right ) 
 \right \rangle , 
 \quad
 \bm{E} \left  ( \bm{r}_j \right ) = 
\pmatrix{E_x \left ( \bm{r}_j \right ) \cr E_y \left ( \bm{r}_j \right )}
 \end{equation}
with $\dagger$ representing Hermitian conjugation.  Moreover, we can collect the four components 
into a single four-dimensional vector $\bm{E}$ with components $E_1 = E_x \left  ( \bm{r}_1 \right )$, 
$E_2 = E_y \left  ( \bm{r}_1 \right )$, $E_3 = E_x \left  ( \bm{r}_2 \right )$, $E_4 = E_y \left  ( \bm{r}_2 
\right )$, so that $\Gamma_{j,k}=  \left \langle E_j E^\ast_k \right \rangle$, for $j,k=1,\ldots,4$. Under 
these conditions every account of coherence and polarization must be a function of these matrices 
$\bm{\Gamma}$ and  $\bm{\Gamma}_{j,k}$.  

\bigskip

\textbf{\textit{Scalar interferometric coherence.--}}
For completeness we recall the standard measure of coherence $\mu$ in the simplest case of two 
scalar fields $E_{1,2}$ 
\begin{equation}
\label{musc}
\mu =  \frac{\left \langle E_1E^\ast_2 \right \rangle}{\sqrt{\left \langle | E_1 |^2 \right \rangle 
\left \langle | E_2 |^2 \right \rangle}} .
\end{equation}
We refer to this as interferometric coherence in the sense of being the key factor controlling the 
visibility of the interference fringes obtained when superimposing $E_{1,2}$.

\bigskip

\textbf{\textit{Polarization.--}}
Polarization expresses the coherence between two components, say $E_{x,y}$ of a transverse field 
vector at a given spacial point as  
\begin{equation}
\label{dP}
P^2 = 2 \frac{\mathrm{tr} \left ( \bm{\Gamma}^2 \right ) }{\left (  \mathrm{tr} \bm{\Gamma} \right )^2 } - 1 ,
\end{equation}
where here $\bm{\Gamma}$ refers to the $2 \times 2$ coherency matrix with matrix elements 
$\Gamma_{j,k} = \left \langle E_j E^\ast_k \right \rangle$, $j,k=x,y$. It is wort noting that this is the 
maximum interferometric coherence (\ref{musc}) that can be reached between any two field components  
$E_{1,2}$  obtained from  $E_{x,y}$ by a unitary $2 \times 2$ matrix $U \in$ U(2), this is to say $\mu (U) 
\leq P$. After computing $P$ for $\bm{E} \left  ( \bm{r}_{1,2} \right )$ the two degrees of polarization 
$P_{1,2}$ can be be combined to provide a single value, for example as in Refs. \cite{CP}.

\bigskip

\textbf{\textit{Vectorial interferometric coherence.--}}
The interferometric account of coherence $\mu$ becomes more complex when the interfering fields 
are vectorial. Accordingly, different definitions have been proposed as different generalizations of 
Eq. (\ref{musc}), such as \cite{KW, TSF,RG}:
\begin{equation}
\label{mW}
\mu_{KW} =  
\frac{\mathrm{tr} \bm{\Gamma}_{1,2} }{\sqrt{\mathrm{tr} \bm{\Gamma}_{1,1} \mathrm{tr} \bm{\Gamma}_{2,2}}},
\end{equation}
\begin{equation}
\label{mW}
\mu^2_{TSF} =  
\frac{\mathrm{tr} \left (  \bm{\Gamma}_{1,2}  \bm{\Gamma}_{1,2}^\dagger \right )}{\mathrm{tr} 
\bm{\Gamma}_{1,1} \mathrm{tr} \bm{\Gamma}_{2,2}},
\end{equation} 
and, when the corresponding inverses exist,
\begin{equation}
\label{mSI}
\mu_{S,I} = \mathrm{singular \; values \; of \; } 
\bm{\Gamma}_{1,1}^{-1/2} \bm{\Gamma}_{1,2} \bm{\Gamma}_{2,2}^{-1/2},
\end{equation} 
where $\mu_{S,I}$ are real and $ \mu_S  \geq \mu_I \geq 0$.

\bigskip

\textbf{\textit{Global coherence.--}}
In this work we are mostly interested in regarding the four field components as a whole, asking for 
the global coherence conveyed by the complete field. This should comprise both the polarization 
and interferometric contributions, and should be expressed by the whole $\bm{\Gamma}$ instead 
of its sub-matrices  $\bm{\Gamma}_{j,k}$. 

For another perspective, we may say that polarization and interferometric coherence depends on 
a choice of field modes, which is the equivalent of the basis dependence in quantum mechanics. 
Thus we can attempt a mode-independent approach where coherence may be referred to as 
intrinsic, {\em per se},  or global \cite{BCP}. This would be analogous to the role played by the 
degree of polarization (\ref{dP}) versus the degree of coherence (\ref{musc}) regarding two scalar 
electric fields.

A convenient generalization of the degree of polarization  (\ref{dP}) to four-dimensional fields can be
\begin{equation}
\label{dPv}
\mu_g^2 = \frac{4}{3} \frac{\mathrm{tr} \left ( \bm{\Gamma}^2 \right ) }{\left (  \mathrm{tr} \bm{\Gamma} 
\right )^2 } - \frac{1}{3} =  \frac{4}{3} \bm{\lambda}^2 - \frac{1}{3} ,
\end{equation}
where  $\bm{\lambda}$ is a four-dimensional vector containing the eigenvalues of  $\bm{\Gamma}/ 
\mathrm{tr} \bm{\Gamma}$ \cite{LU}. An alternative assessment along this line can be found in 
Ref. \cite{OYK}. In the next section we show that $\mu_g$ is actually a case of  R\'{e}nyi entropy.

\section{Majorization and global degree of coherence}

Next we recall the idea of majorization and its relation with coherence measures as functions of the 
$N \times N$ coherency matrix $\bm{\Gamma}$. The main idea is that coherence is reflected in the 
dispersion of the eigenvalues of $\bm{\Gamma}/\mathrm{tr} \bm{\Gamma}$, that are real, positive, 
and normalized $\sum_{j=1}^N \lambda_j = 1$.  We have two clear extremes. We have full coherence 
for those $\bm{\Gamma}_p$  with only one eigenvalue different from zero, say $\lambda_1 = 1, 
\lambda_{j \neq 1} =0$. On the other extreme, there is complete lack of coherence for those 
$\bm{\Gamma}_I$ where all the eigenvalues are equal $\lambda_j = 1/N$, so that $\bm{\Gamma}_I$ 
is proportional to the identity. In between, the degree of coherence may be assessed using many 
possible functions of $\bm{\lambda}$. Most of them are of entropic nature, such as the R\'{e}nyi 
entropies \cite{RB}
\begin{equation}
\label{R}
R_q (\bm{\lambda})= \frac{1}{1-q} \ln \left ( \sum_{j=1}^N \lambda_j^q\right ),
\end{equation}
where $q > 0$ is an index labeling different entropies. The limiting case $q \rightarrow 1$  is the 
Shannon entropy $R_1 =  -  \sum_{j=1}^N  \lambda_j \ln \lambda_j$.  For example, we have that  
$\mu_g $ in Eq. (\ref{dPv}) is essentially $R_2$ as 
\begin{equation}
\mu_g^2 = \frac{4}{3}e^{-R_2} -\frac{1}{3} .
\end{equation}

\begin{figure}
\centering
\includegraphics[width=5cm]{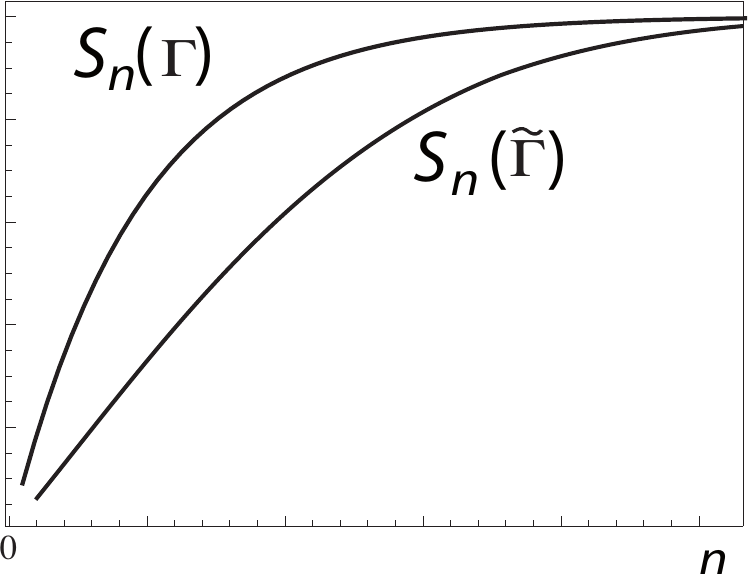}
\caption{Relation between partial ordered sums $S_n$ when the majorization  $\tilde{\bm{\Gamma}} 
\prec \bm{\Gamma}$ holds.}
\end{figure}

We say that $\bm{\Gamma}$ majorizes  $\tilde{\bm{\Gamma}}$, which will be expressed as 
$\tilde{\bm{\Gamma}} \prec \bm{\Gamma}$, when the following relation between all the ordered 
partial sums $S_n$ of their corresponding eigenvalues holds, 
 \begin{equation}
 \label{major}
S_n  \left ( \tilde{\bm{\Gamma}} \right ) =  \sum_{j=1}^n  \tilde{\bm{\lambda}}^{\downarrow} _j \leq 
\sum_{k=1}^n \bm{\lambda}^{\downarrow} _k = S_n  \left ( \bm{\Gamma} \right  )  ,
 \end{equation}
for all $n=1, 2 \ldots N$, where the superscript $\downarrow$  denotes the same  $\bm{\lambda}_j$ 
but arranged in decreasing  order
\begin{equation}
\lambda^{\downarrow} _1 \geq \lambda^{\downarrow} _2 \geq \ldots \geq \lambda^{\downarrow} _N .
\end{equation}
Throughout we will say that $\bm{\Gamma}$ and $\tilde{\bm{\Gamma}}$ are comparable if one 
majorizes the other. 

\bigskip

Next, some interesting facts about majorization. Majorization holds if and only if  $\tilde{\bm{\lambda}}
= M {\bm{ \lambda}}$ where $M$ is a doubly stochastic matrix, so that $\tilde{\bm{\Gamma}}$ is more 
uniform or more mixed than $\bm{\Gamma}$. For all $\bm{\Gamma}$ we have $\bm{\Gamma}_I \prec 
\bm{\Gamma} \prec \bm{\Gamma}_p$ \cite{BZ}. Whenever two $\bm{\Gamma}$ are comparable, the 
result is respected by all Schur-concave functions that includes the R\'{e}nyi entropies: if  $\tilde{\bm{\Gamma}} 
\prec \bm{\Gamma}$ then $R_q (\tilde{\bm{\Gamma}} ) >  R_q (\bm{\Gamma} )$ for all $q$. 

\bigskip

After all these facts we may say if $\tilde{\bm{\Gamma}} \prec \bm{\Gamma}$ then $\bm{\Gamma}$ 
is more coherent than  $\tilde{\bm{\Gamma}}$. Majorization is a partial ordering relation, so that there 
are incomparable states: this is neither $\tilde{\bm{\Gamma}} \prec \bm{\Gamma}$ nor $\bm{\Gamma} 
\prec \tilde{\bm{\Gamma}}$. In such a case the  ordered sums in Fig. 1 will intersect and the entropies 
will provide contradictory conclusions, such that  $R_q (\tilde{\bm{\Gamma}} ) > R_q (\bm{\Gamma})$ 
while  $R_{p} (\tilde{\bm{\Gamma}} ) < R_{p} ( \bm{\Gamma})$ for different entropies $p \neq q$.

\bigskip

The case of $2 \times 2$ matrices $\bm{\Gamma}$ is rather trivial since after normalization the spectrum 
$\bm{\lambda}$ depends on a single parameter. Thus any two $\bm{\Gamma}$ are comparable and 
there is no room for ambiguities nor discrepancies between measures.  The case of $3 \times 3$ 
matrices $\bm{\Gamma}$ has been completely addressed in a recent work \cite{GJ} regarding  
$\bm{\Gamma}$ as representing polarization in three dimensions. Thus in this work we address  
$4 \times 4$ matrices $\bm{\Gamma}$, that may be representing the two transversal fields described 
above. In the next sections we analyze some relevant and illustrative cases.

\section{Unpolarized beams of the same intensity}

In this case by means  of suitable U(2) transformations the $\bm{\Gamma}$ matrix can arranged so that
\begin{equation}
\label{up}
\bm{\Gamma}_{1,1} = \bm{\Gamma}_{2,2} = I \pmatrix{1 & 0 \cr 0 & 1}, 
\quad
\bm{\Gamma}_{1,2} = I \pmatrix{\mu_S & 0 \cr 0 & \mu_I},   
\end{equation}
where $I$ represents the intensity of each component, that will play no role on the following. 

\bigskip

The corresponding values of the above measures of polarization and interferometric coherence are: 
$P_1 = P_2 =0$,
\begin{equation}
\label{muv1}
\mu_{KW} = \frac{1}{2} \left ( \mu_S + \mu_I  \right ) , \quad
\mu_{TSF}^2 = \frac{1}{4} \left ( \mu_S^2 + \mu_I^2  \right ) ,  
\end{equation}
while the global coherence is 
\begin{equation}
\label{mg}
\mu_g^2 = \frac{1}{6}  \left ( \mu_S^2 + \mu_I^2 \right ) .
\end{equation}
The properly ordered eigenvalues of $\bm{\Gamma}/\mathrm{tr} \bm{\Gamma}$  in decreasing order 
are
\begin{eqnarray}
& \lambda_1^\downarrow = \frac{1}{4} \left ( 1 + \mu_S \right ),  \quad  
\lambda_2^\downarrow = \frac{1}{4} \left ( 1 + \mu_I \right ), & \nonumber \\
& \lambda_3^\downarrow = \frac{1}{4} \left ( 1 - \mu_I \right ),  \quad  
\lambda_4^\downarrow = \frac{1}{4} \left ( 1 - \mu_S \right ), &
\end{eqnarray}
leading to the following ordered partial sums
\begin{eqnarray}
S_1 & = & \frac{1}{4} \left ( 1 + \mu_S \right ),  \quad  
S_2 = \frac{1}{4} \left ( 2 + \mu_S + \mu_I \right ),  \nonumber \\
S_3 & = & \frac{1}{4} \left ( 3  + \mu_S \right ),  \quad  
S_4  =1.  
\end{eqnarray}
After these expressions for $S_n$ majorization is actually determined just by the two first conditions 
in Eq. (\ref{major}) so that  $\tilde{\bm{\Gamma}} \prec \bm{\Gamma}$ if and only if the following two 
conditions are satisfied 
\begin{equation}
\label{mc}
\mu_S \geq  \tilde{\mu}_S, \quad 
\mu_S + \mu_I \geq  \tilde{\mu}_S + \tilde{\mu}_I .
\end{equation}

Since the degree of polarization of $\bm{E} \left  ( \bm{r}_j \right )$  vanishes one might ask 
whether the majorization conditions (\ref{mc}) are equivalent to any of the above degrees of 
interferometric coherence in Eq.  (\ref{muv1}) or the global coherence in Eq. (\ref{mg}). The 
result is negative. To show this the picture in Fig. 2 may be useful. This is a $\mu_{S,I}$ plane 
where any $\bm{\Gamma}$ is represented by  a point in the region of the first quadrant  below 
the bisecting line. This is to respect the condition $\mu_S \geq \mu_I \geq 0$. The dotted line 
represents the condition $\mu_S \geq  \tilde{\mu}_S$, while the dashed line represents the 
condition $\mu_S + \mu_I \geq  \tilde{\mu}_S + \tilde{\mu}_I $.  Therefore relations (\ref{mc}) 
define three regions: 

$\alpha$: Points $\tilde{\bm{\Gamma}}$ with $\tilde{\bm{\Gamma}} \prec \bm{\Gamma}$ 
are to the left of the dotted line and below the dashed line.

$\beta$: Points $\tilde{\bm{\Gamma}}$ with $\bm{\Gamma} \prec \tilde{\bm{\Gamma}} $ 
are to the right of the dotted line and above the dashed line.

$\gamma$:  Points $\tilde{\bm{\Gamma}}$ incomparable with $\bm{\Gamma}$.

After this it is clear from Eqs. (\ref{muv1}) and  (\ref{mc}) that $\mu_{KW} \geq  \tilde{\mu}_{KW}$ 
is just a necessary but not sufficient condition for  $\tilde{\bm{\Gamma}} \prec \bm{\Gamma}$.  
On the other hand, both $\mu_{TSF}$  and $\mu_g$ depend just on the distance of the point 
$\bm{\Gamma}$ to the origin. Some simple geometry shows that within the region $\mu_S \geq 
\mu_I$ the circle passing through $\bm{\Gamma}$ lies always on the $\gamma$ sectors, so 
that  $\mu_{TSF}$ and $\mu_g$ provide necessary but not sufficient conditions regarding 
majorization. 

\bigskip

A simple example of incomparable states is provided by  $\mu_S =1$, $\mu_I = 0$ 
and $1 \geq \tilde{\mu}_S = \tilde{\mu}_I \geq 1/2$. In this case we have always $\tilde{\mu}_{KW} 
\geq \mu_{KW}$. In Fig. 3 we have represented $R_2 (\tilde{\bm{\Gamma}}) - R_2 (\bm{\Gamma})$ 
and $R_1 (\tilde{\bm{\Gamma}}) - R_1 (\bm{\Gamma})$. It can be appreciated that for $0.78 >  
\tilde{\mu}_S >0.71$ the two R\'{e}nyi entropies provide contradictory conclusions: this is  
$R_1 (\tilde{\bm{\Gamma}}) > R_1 (\bm{\Gamma})$ versus $R_2 (\tilde{\bm{\Gamma}}) < 
R_2 (\bm{\Gamma})$. We have plotted as well $\tilde{\mu}^2_{TSF} - \mu^2_{TSF}$ around 
this same region showing regions where this measure of interferometric coherence contradicts 
the entropies. Moreover, in Fig. 4 we plot the ordered partial sums $S_n (\bm{\Gamma})$ 
and $S_n (\tilde{\bm{\Gamma}})$ for $ \tilde{\mu}_S = \tilde{\mu}_I = 0.75$ showing the lack 
of majorization.

\begin{figure}[top]
\begin{center}
\includegraphics[width=5cm]{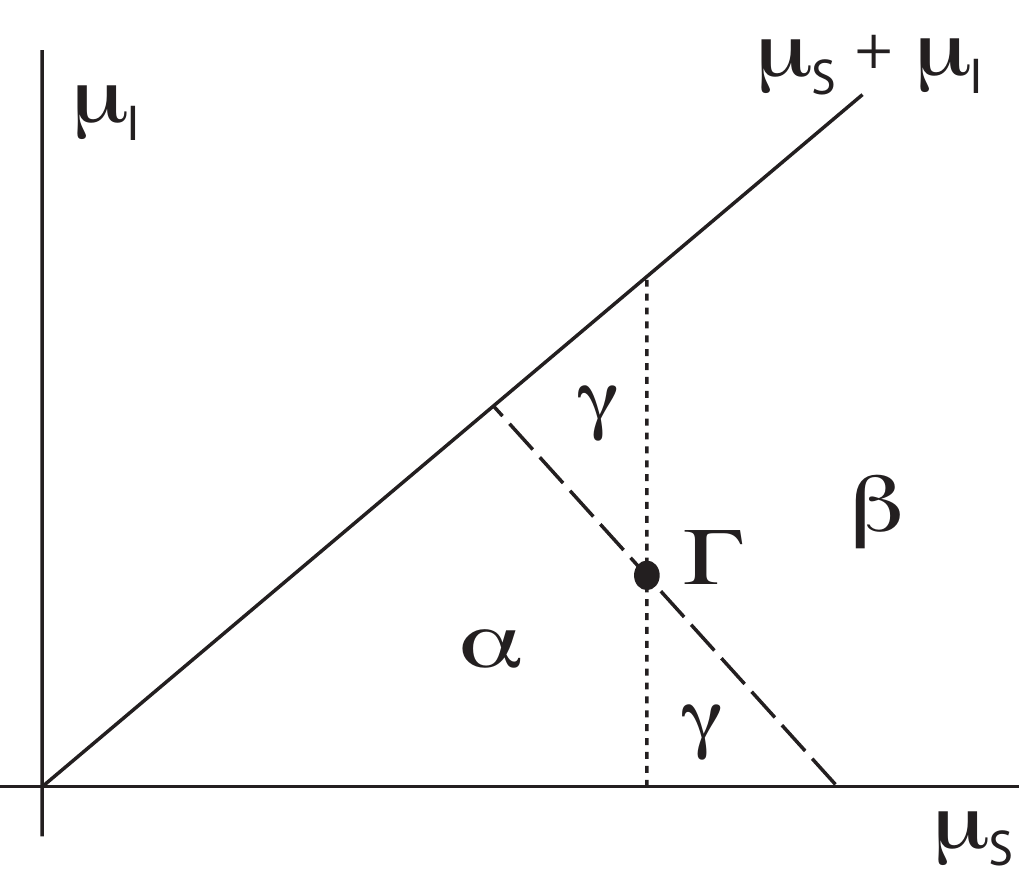}
\end{center}
\caption{Graphical translation of relations (\ref{mc}). The dotted line represents the condition 
$\mu_S \geq  \tilde{\mu}_S$, while the dashed line represents the second condition $\mu_S 
+ \mu_I \geq  \tilde{\mu}_S + \tilde{\mu}_I $.}
\end{figure}

\begin{figure}[top]
\begin{center}
\includegraphics[width=7cm]{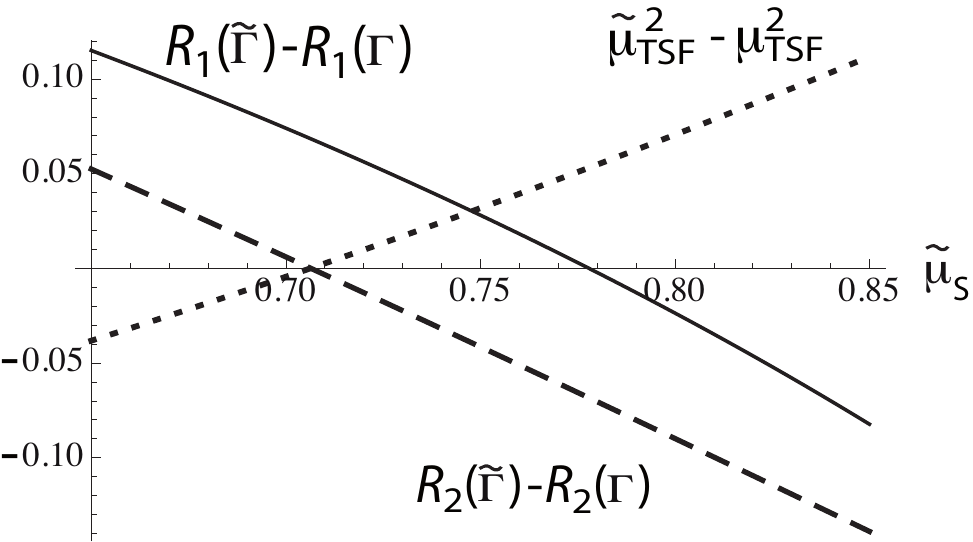}
\end{center}
\caption{Plot of $R_1 (\tilde{\bm{\Gamma}}) - R_1 (\bm{\Gamma})$ (solid) and $R_2 (\tilde{\bm{\Gamma}}) - 
R_2 (\bm{\Gamma})$ (dashed), and $\tilde{\mu}^2_{TSF} - \mu^2_{TSF}$ (dotted), as functions of $\tilde{\mu}_S$ 
for $\mu_S =1$, $\mu_I = 0$ and $\tilde{\mu}_S = \tilde{\mu}_I  $ .}
\end{figure}

\begin{figure}[top]
\begin{center}
\includegraphics[width=6cm]{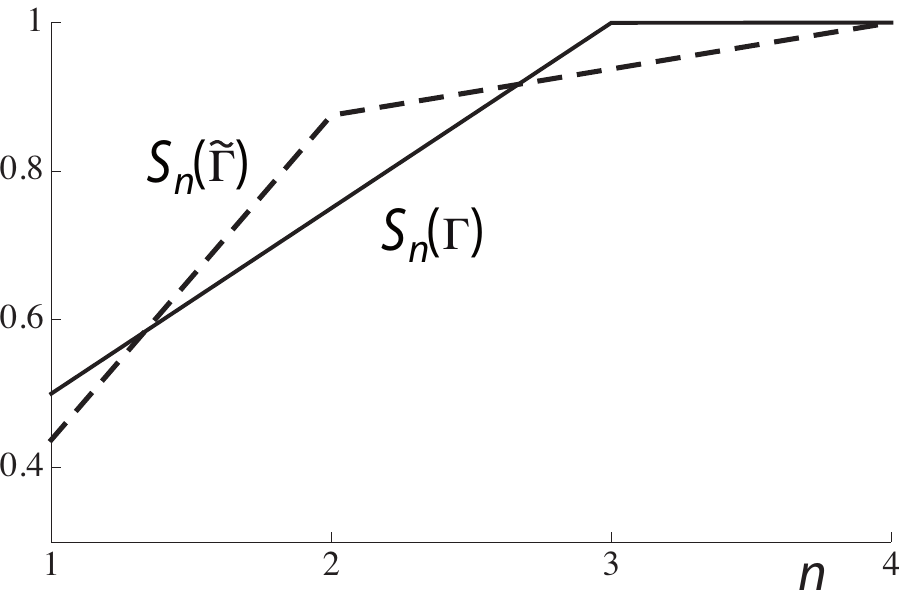}
\end{center}
\caption{Plot of the ordered partial sums $S_n (\bm{\Gamma})$ (solid) and $S_n (\tilde{\bm{\Gamma}})$
(dashed)  for two unpolarized waves with $\mu_S =1$, $\mu_I = 0$, and $ \tilde{\mu}_S = \tilde{\mu}_I =0.75$,
showing the lack of majorization. Points have been joined by continuous lines as an aid to the eye.}
\end{figure}

\section{Fully polarized versus partially polarized beams}

Let us combine a fully polarized beam linearly polarized along then axis $x$ with a partially polarized 
beam with 
\begin{eqnarray}
& \bm{\Gamma}_{1,1} = I \pmatrix{1 & 0 \cr 0 & 0}, 
\quad
\bm{\Gamma}_{2,2} = I \pmatrix{1 & 0 \cr 0 & \delta},  & \nonumber \\
& \bm{\Gamma}_{1,2} = I \pmatrix{\mu_S & 0 \cr 0 & 0},   &
\end{eqnarray}
where we will assume $\delta \geq 0$, and $I$ represents the intensity of each component that will 
play no role on the following. 

The corresponding values of the measures of polarization and interferometric coherence are:
\begin{equation}
\label{idc}
P_1 = 1, \quad P_2 =\frac{|1-\delta |}{1+\delta}, 
\end{equation}
\begin{equation}
\mu_{TSF}^2 =\mu_{KW}^2 = \frac{\mu_S^2}{1+\delta} ,
\end{equation}
and 
\begin{equation}
 \mu_g^2 = \frac{4 + 8 \mu_S^2 + 3 \delta^2 - 4 \delta}{3 \left ( 2 + \delta \right )^2} .
\end{equation}
The dependence on $\mu_S$ is the expected one: larger $\mu_S$  implies both larger interferometric 
coherence and larger global coherence. On the other hand, the dependence on $\delta$ is more 
complicated. The degree of polarization has a minimum at  $\delta = 1$, while interferometric coherence 
always  decreases when increasing $\delta$. Regarding global coherence we have that for fixed 
$\mu_S$ it has a minimum at $\delta = 1+ \mu_S^2$. Thus this example provides an interesting 
competition between polarization and interferometric coherence.  From now on we will focus on the 
dependence on $\delta$ for fixed $\mu_S$. 

\bigskip

In this case, since there is always a vanishing  eigenvalue we have $S_3 = S_4 = 1$ and majorization 
is just determined by the first two partial sums in Eq. (\ref{major}). Taking into account that 
$S_1 = \lambda^\downarrow_1$ and $S_3 =  S_2 + \lambda^\downarrow_3 = 1$ we get a very 
simple relation already found in the three-dimensional problem considered in Ref. \cite{GJ}. This 
is that $\tilde{\bm{\Gamma}} \prec \bm{\Gamma}$ is equivalent to
\begin{equation}
\label{gj}
\lambda_1^\downarrow  \geq \tilde{\lambda}_1^\downarrow , \quad
\lambda_3^\downarrow  \leq \tilde{\lambda}_3^\downarrow .
\end{equation}

Regarding the ordering of the eigenvalues $\bm{\lambda}$,  we can distinguish three cases depending 
on the relation between $\delta$ and $\mu_S$. The first one we consider is $\delta < 1 -\mu_S$, where 
the arrangement of $\bm{\lambda}$ in decreasing order is 
\begin{eqnarray}
\label{fo}
\lambda_1^\downarrow & =  & \frac{1 + \mu_S}{2 + \delta},  \quad  
\lambda_2^\downarrow = \frac{1 - \mu_S}{2 + \delta},  \nonumber \\
 \lambda_3^\downarrow & = & \frac{\delta}{2 + \delta},  \quad  
\lambda_4^\downarrow = 0 .
\end{eqnarray}
After Eqs. (\ref{gj}) and (\ref{fo})  for $\mu_S = \tilde{\mu}_S$ we readily get  
\begin{equation}
\tilde{\bm{\Gamma}} \prec \bm{\Gamma} \leftrightarrow \tilde{\delta} \geq  \delta .
\end{equation}
This is a quite expected result since for $\delta  <1$ increasing $\delta$ means both lesser degree 
of polarization and lesser interferometric coherence. 

\bigskip

The opposite situation holds for $\delta > 1  + \mu_S$. In this case we will have $\delta >1$  so we may 
expect that global coherence should emerge of a suitable balance between the opposed behaviors 
displayed by the degree of polarization and the interferometric coherence. The ordering of eigenvalues is
\begin{eqnarray}
\label{so}
\lambda_1^\downarrow &  = &  \frac{\delta}{2 + \delta},  \quad  
\lambda_2^\downarrow = \frac{1 + \mu_S}{2 + \delta},  \nonumber \\
\lambda_3^\downarrow &  =  &  \frac{1 - \mu_S}{2 + \delta} ,  \quad  
\lambda_4^\downarrow = 0 .
\end{eqnarray}
After Eqs. (\ref{gj}) and (\ref{so})  for $\mu_S = \tilde{\mu}_S$ we readily get 
\begin{equation}
\tilde{\bm{\Gamma}} \prec \bm{\Gamma} \leftrightarrow \tilde{\delta} \leq \delta .
\end{equation}
Roughly speaking, when $\delta$ increases the purity of the state provided by polarization overwhelms  
the decrease of the interferometric coherence.

\bigskip

Finally,  in the intermediate situation $1  + \mu_S > \delta > 1  - \mu_S $  and $\mu_S = \tilde{\mu}_S$ 
the states are always incomparable unless $\tilde{\delta} = \delta$. This is because 
\begin{eqnarray}
\lambda_1^\downarrow &  = &  \frac{1 + \mu_S}{2 + \delta},  \quad  
\lambda_2^\downarrow =  \frac{\delta}{2 + \delta},  \nonumber \\
\lambda_3^\downarrow &  =  &  \frac{1 - \mu_S}{2 + \delta} ,  \quad  
\lambda_4^\downarrow = 0 ,
\end{eqnarray}
so that the two conditions in Eq. (\ref{gj}) are incompatible
\begin{equation}
\lambda_1^\downarrow  \geq \tilde{\lambda}_1^\downarrow \leftrightarrow  \tilde{\delta} \geq \delta  ,
\quad \lambda_3^\downarrow  \leq \tilde{\lambda}^\downarrow_3   \leftrightarrow  \tilde{\delta} \leq  
\delta .
\end{equation}

\section{Conclusions}

We have shown that majorization provides a powerful approach to the coherence conveyed by partially 
polarized  transversal waves. This is because it can be regarded as a kind of meta-measure of global 
coherence whose conclusions are respected by entropic measures of polarization and coherence. 
Moreover, majorization allows us to draw many parallels with coherence in quantum physics.
 
We have illustrated the approach by means of some  simple but  meaningful examples. The results are 
contrasted to other measures of polarization and interferometric coherence for vectorial waves. The 
situation is particularly interesting when  polarization and interference behave in opposite ways.

\section*{ACKNOWLEDGMENTS}

A. L. acknowledges support from Project No. FIS2012-35583
of the Spanish Direcci\'{o}n General de Investigaci\'{o}n del 
Ministerio de Econom\'{\i}a y Competitividad, and from 
Comunidad Aut\'{o}noma de Madrid research  consortium QUITEMAD+ 
grant S2013/ICE-2801.


\begin{thebibliography}{00}

\bibitem{BCP}
T. Baumgratz, M. Cramer, and M. B. Plenio, Quantifying coherence, Phys. Rev. Lett. {\bf 113}, 140401 (2014);
S. Cheng and M. J. W. Hall, Complementarity relations for quantum coherence, Phys. Rev. A {\bf 92}, 042101 (2015).

\bibitem{KW}
B. Karczewski,  Degree of coherence of the electromagnetic field, Phys. Lett. {\bf 5}, 191--192 (1963);
E. Wolf, Unified theory of coherence and polarization of random  electromagnetic beams, Phys. Lett. A {\bf 312}, 263--267 (2003).

\bibitem{TSF}
J. Tervo, T. Set\"{a}l\"{a}, and A. T. Friberg, Degree of coherence for electromagnetic fields, Opt. Express {\bf 11}, 1137--1143 (2003).

\bibitem{RG}
P. R\'{e}fr\'{e}gier and F. Goudail, Invariant degrees of coherence of partially polarized light, Opt. Express {\bf 13}, 6051--6060 (2005).

\bibitem{OYK}
H. M. Ozaktas, S. Y\"{u}ksel, and M. A. Kutay, Linear algebraic theory of partial coherence: discrete fields and 
measures of partial coherence, J. Opt. Soc. Am. A {\bf 19}, 1563--1571 (2002).

\bibitem{LU}
A. Luis, Degree of coherence for vectorial electromagnetic fields as the distance between correlation matrices, J. Opt. Soc. Am. A {\bf 24}, 1063--1068 (2007);
A. Luis, Overall degree of coherence for vectorial electromagnetic fields and the Wigner function, J. Opt. Soc. Am. A {\bf 24}, 2070--2074 (2007);
A. Luis, Coherence and visibility for vectorial light, J. Opt. Soc. A {\bf 27}, 1764--1769 (2010).

\bibitem{QvC1}
A. Luis, Quantum-classical correspondence for visibility, coherence, and relative phase for multidimensional systems, Phys. Rev. A {\bf 78}, 025802 (2008).

\bibitem{QvC2}
O. Gamel and D. F. V. James, Measures of quantum state purity and classical degree of polarization Phys. Rev. A {\bf 86}, 033830 (2012).

\bibitem{PP}
J. J. Gil, Polarimetric characterization of light and media. Physical quantities involved in polarimetric phenomena, Eur. Phys. J. Appl. Phys. {\bf 40}, 1Ð47 (2007);
J. J. Gil, Interpretation of the coherency matrix for three-dimensional polarization states, Phys. Rev. A {\bf 90}, 043858 (2014);
J. J. Gil and R. Ossikovski, {\it Polarized Light and the Mueller Matrix Approach} (CRC Press, 2016). 

\bibitem{PE} 
Ch. Brosseau, \textit{Fundamentals of Polarized Light: A Statistical Optics Approach} (Wiley,  1998);
A. Luis, Degree of polarization in quantum optics, Phys. Rev. A \textbf{66}, 013806 (2002);
A. Picozzi, Entropy and degree of polarization for nonlinear optical waves, Opt. Lett. \textbf{29}, 1653--1655 (2004);
Ph. R\'{e}fr\'{e}gier, Polarization degree of optical waves with non-Gaussian probability density 
functions: Kullback relative entropy-based approach, Opt. Lett. \textbf{30}, 1090--1092 (2005);
Ph. R\'{e}fr\'{e}gier and F. Goudail, Kullback relative entropy and characterization of partially polarized 
optical waves,  J. Opt. Soc. Am. A {\bf 23}, 671--678 (2006).

\bibitem{MO}
A. W. Marshall and I. Olkin,  {\it Inequalities: Theory of Majorization and Its Applications}  (Academic Press, New York, 1980);
J. Aczel and Z. Daroczy, {\it On Measures of Information and Their Characterization} (Academic Press, New York, 1975). 

\bibitem{RB}
A. R\'enyi, On the measures of entropy and information, Proc. 4th Berkeley Symp. on Mathematics and Statistical Probability (University of California Press,
1961), Vol. 1, pp. 547--561;
Ch. Beck, Generalised information and entropy measures in physics, Contemporary Physics \textbf{50}, 495--510 (2009).

\bibitem{GJ}
O. Gamel and D. F. V. James, Majorization and measures of classical polarization in three dimensions,
J. Opt. Soc. Am. A {\bf 31}, 1620--1626 (2014).

\bibitem{CP} 
G. Piquero, J. M. Movilla, P. M. Mej\'{\i}as, R. Mart\'{\i}nez-herrero, Degree of polarization of non-uniformly partially polarized beams: a proposal,
Opt. Quantum Electron. {\bf 31}, 223--226 (1999);
A. Luis, Coherence, polarization, and entanglement for classical light fields, Opt. Commun. {\bf  282}, 3665--3670 (2009).

\bibitem{BZ}
I. Bengtsson and K. \.{Z}yczkowski, Geometry of Quantum States: An Introduction to Quantum Entanglement, Cambridge University Press, Cambridge, 2006.

\end{thebibliography}
\end{document}